\documentclass[twocolumn,showpacs,preprintnumbers,amsmath,amssymb,superscriptaddress]{revtex4}

\usepackage[dvips]{color,graphicx}

\usepackage{graphicx}

\begin{document}
\draft

\title{Interface Magnetoresistance in Manganite-Titanate Heterojunctions}

\author{T. Susaki}
\affiliation{Department of Advanced Materials Science, University of Tokyo, Chiba 277-8561, Japan}

\author{N. Nakagawa$^*$}
\affiliation{Department of Advanced Materials Science, University of Tokyo, Chiba 277-8561, Japan}
\affiliation{Japan Science and Technology Agency, Kawaguchi, 332-0012, Japan}

\author{H. Y. Hwang}
\affiliation{Department of Advanced Materials Science, University of Tokyo, Chiba 277-8561, Japan}
\affiliation{Japan Science and Technology Agency, Kawaguchi, 332-0012, Japan}

%\author{T. Susaki,$^{1}$\, N. Nakagawa,$^{1,2,*}$ and H. Y. Hwang$^{1,2}$}
%\address{Department of Advanced Materials Science, University of Tokyo, Chiba 277-8561, Japan}
%\address{Japan Science and Technology Agency, Kawaguchi, 332-0012, Japan}

\date{\today} 

\begin{abstract}
We have found that the current-voltage characteristics of La$_{0.7}$Sr$_{0.3}$MnO$_{3(-\delta)}$/Nb:SrTiO$_3$ rectifying junctions are quantitatively well-described by (thermally-assisted) tunneling with an effectively temperature-independent Schottky barrier under no magnetic field, while those of the oxygen deficient junction remarkably deviate from such a simple behavior as magnetic field is applied. These results indicate a new form of magnetoresistance arising from magnetic field changes of the interface band diagram via the strong electron-spin coupling in manganites. 
%(587 characters)
\end{abstract}
\pacs{PACS numbers: 73.20.-r,73.30.+y, 75.70.Cn}
\maketitle

%\begin{multicols}{2}
%\narrowtext
The large magnetotransport effect of manganites both in bulk and in layered structures is one of the outstanding phenomena observed in transition-metal oxides.\cite{Tokura} In double-exchange manganites, external magnetic field aligns the fluctuating $t_{2g}$ spins near the Curie temperature ($T_C$), significantly increasing the conductivity of  the $e_g$ electron due to a strong Hund's-rule ferromagnetic coupling between the $e_g$ electron spin and the $t_{2g}$ local spin (colossal magnetoresistance, CMR). On the other hand,  in manganite-insulator-manganite spin tunnel junctions,\cite{JZSun} or at polycrystalline grain boundaries,\cite{GB} external magnetic field induces a switching between
a high resistance and low resistance state,  which corresponds to the switching between ferromagnetic and antiferromagnetic alignment of the manganite electrodes or grains (tunneling magnetoresistance, TMR). A large change in the resistance in TMR originates in 
 the 100 \% spin-polarized character of the valence band \cite{Park} and hence is significant at lower temperature, while CMR appears near $T_C$.
Very recently, however, it has been reported that some single-interface manganite-titanate  junctions, 
which are the most simple building blocks for artificial manganite structures, show
significant junction magnetoresistance.\cite{Sun_November,Lu,Jin,Nakagawa} In particular, junctions formed with La$_{0.7}$Sr$_{0.3}$MnO$_{3-\delta}$ exhibited large magnetoresistance and magnetocapacitance, unlike those formed with La$_{0.7}$Sr$_{0.3}$MnO$_3$.\cite{Nakagawa} Since the observed
junction magnetoresistance occurs with no spin filter, the mechanism
for such junction magnetoresistance must be different from either CMR or TMR. 

In this paper we analyze the temperature dependence of junction current-voltage ($I-V$) characteristics of single-interface manganite-titanate junctions with and without junction magnetoresistance. Junction $I-V$ measurement is a powerful technique to
study interface electronic structure since: (1) it can be performed even under external magnetic field, and
(2) it sensitively probes the very interface. This magnetic-field
compatibility is absent in various electron spectroscopy techniques, all of which are quite useful
to probe the electronic structure of transition metal oxides when no magnetic field is applied ($H=0$).\cite{spin_resolved} 
At $H=0$, the temperature-dependent $I-V$ characteristics of the manganite-titanate junctions studied here are fully consistent with  the (thermally-assisted) tunneling model. 
(Thermally-assisted) tunneling is relevant when the Schottky barrier is relatively thin. Since the carrier concentration is generally large and the barrier width is correspondingly thin in transition-metal oxide junctions, the present analysis  should be quite generally relevant for understanding oxide junction characteristics.
However, the $I-V$ characteristics of  the magnetoresistive La$_{0.7}$Sr$_{0.3}$MnO$_{3-\delta}$ junction are quite unusual for $H \ne 0$, stongly deviating from conventional thermally assisted tunneling. 
We find that the magnetoresistance arises from changes of the Schottky barrier by magnetic field, originating from double exchange coupling to the electronic structure of the interface.

We deposited oxygen deficient La$_{0.7}$Sr$_{0.3}$MnO$_{3-\delta}$ and oxygen stoichiometric La$_{0.7}$Sr$_{0.3}$MnO$_3$ films with 80 nm thickness on Nb 0.01 wt \% - doped SrTiO$_3$ (001) single-crystal substrates by pulsed laser ablation (KrF excimer laser). Sr and Nb substitution introduces hole and electron doping in LaMnO$_3$ and SrTiO$_3$, respectively. The pulse frequency was
 4 Hz and the laser fluency at the target surface was $\sim$ 3 J/cm$^2$. 
%The substrates were first annealed at 850 $^{\circ}$C for 30 minutes to remove the contamination at the surface and then cooled down to the growth temperature of 700 - 750 $^{\circ}$C. 
The substrate temperature was 700 - 750 $^{\circ}$C and the oxygen partial pressure was 250 mTorr for stoichiometric films and 1 mTorr for oxygen deficient films during the growth. As oxygen deficiencies are introduced, the hole concentration decreases in manganites:\cite{Li} the insulator-metal transition temperature was $\sim$ 180 K for the oxygen deficient film studied here and $\sim$ 350 K for oxygen stoichiometric film. 
 Evaporated Au and Al were used as the Ohmic electrodes for La$_{0.7}$Sr$_{0.3}$MnO$_{3(-\delta)}$ films and for Nb:SrTiO$_3$ substrates. We define forward bias to correspond to positive bias of the manganite, and the direction of magnetic field was perpendicular to the interface plane.
 
\begin{figure}
\begin{center}
\includegraphics[width=7.5cm]{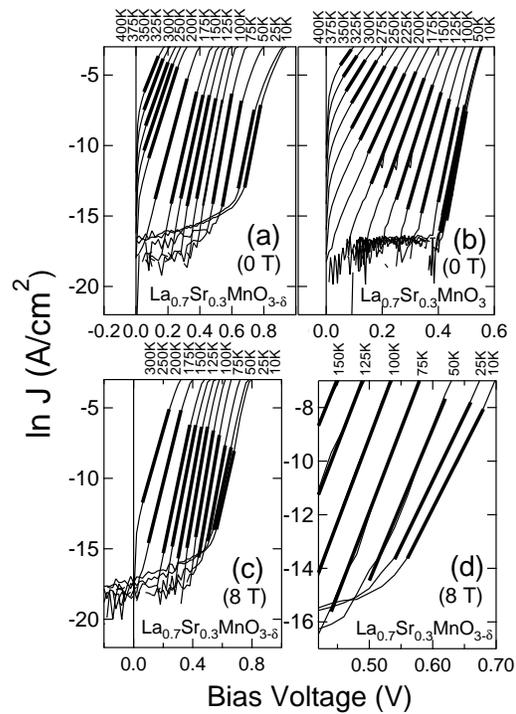}
     \caption{
     Forward-bias $I-V$ characteristics for the oxygen deficient (a) and stoichiometric junction (b) at 0 T. Characteristics for the oxygen deficient junction at 8 T are shown in (c) as well as in (d) on an enlarged scale.  Bold lines are linear fitting on a semilog scale. 
}
\label{fig:IV}
\end{center}
\end{figure}

\begin{figure}
\begin{center}
\includegraphics[width=7.5cm]{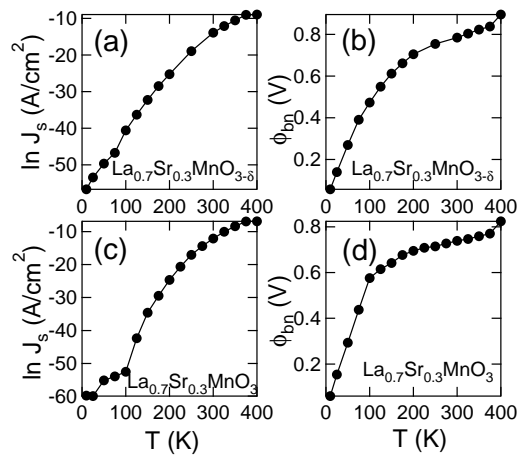}
     \caption{
     Saturation current density $J_S(T)$ (a) and barrier height $\phi_{bn}$ (b) as a function of temperature for the oxygen deficient junction at 0 T. $J_S(T)$ and $\phi_{bn}$ for the stoichiometric junction at 0 T are plotted in panels (c) and (d), respectively. 
          %The magnetic field dependence of $J_S(T)$ and $\phi_{bn}$ are shown in panels (e) and (f).
     }
     \label{fig:JS}
\label{ARPES}
\end{center}
\end{figure}

Figure~\ref{fig:IV} (a) and (b) show the forward-bias $I-V$ characteristics for the oxygen deficient and oxygen stoichiometric manganite-titanate junctions at 0 T.  Here, undoped LaMnO$_3$ is a correlated-electron insulator while undoped SrTiO$_3$ is a
band insulator. Although it is an open question how to model the interface between a doped correlated insulator and
doped band insulator,  since the gap size of undoped LaMnO$_3$ is  much smaller than that of undoped SrTiO$_3$, the present heterojunction may be approximated by the metal-(n-type) semiconductor Schottky junction, where 
only electrons contribute to the junction transport process.\cite{Milnes} The thermionic emission current density (current density $J \equiv I/$(junction area)) across the
Schottky barrier is given by \cite{Sze}
\begin{eqnarray}
J  \simeq  J_S(T){\rm exp}[qV/nkT]
\label{eq:thermionic2}
\end{eqnarray}
for $V \gg kT/q$. Here, $J_S(T)$ is the saturation current density, $q$ is the electronic charge, 
$n$ is the ideality factor (equal to unity in a purely thermionic emission case), 
and $k$ is the Boltzmann constant. 
Panels (a) and (b) of Fig.~\ref{fig:IV} show that the current is linear in the forward bias voltage on a semilog scale and the slope of $I-V$ characteristics increases in going from 400 K to 100 K, qualitatively consistent with thermionic emission current,\cite{Postma,Sawa} while it is almost independent of the temperature below 100 K in both junctions. 
By fitting Eq.~\ref{eq:thermionic2} to the observed $I-V$ characteristics, we determined  $J_S(T)$ and then extracted the temperature dependence of the barrier height $\phi_{bn}$ using the relation \cite{Sze}
\begin{eqnarray}
J_S(T) = A^*T^2{\rm exp}(-\frac{q\phi_{bn}}{kT}),
%\hspace{0.5cm} 
\label{eq:barrier}
\end{eqnarray}
where we set the effective Richardson constant $A^*$ to be 156 A cm$^{-2}$K$^{-2}$, which corresponds to the effective mass $m^*/m_0$
= 1.3 for Nb:SrTiO$_3$.\cite{Sroubek} We show the temperature dependence of $J_S(T)$ and $\phi_{bn}$ for both junctions  at $H=0$ in Fig.~\ref{fig:JS}. 
In both junctions the barrier height $\phi_{bn}$ gradually decreases as the temperature is lowered and is finally reduced to an unphysically small value of $\sim$ 0.05 V at 10 K.

 Temperature-independent slope in the $I-V$ characteristics was previously observed in a more conventional Schottky junction fabricated with Au and heavily doped ($\sim 5 \times 10^{17}$ cm$^{-3}$) GaAs.\cite{Padovani}
The deviation from the ideal thermionic emission was attributed to the (thermally-assisted) tunneling process. 
At low temperature, electrons between the Fermi level ($E_F$) and conduction-band bottom tunnel to the metal under forward bias (direct tunneling), while at higher temperature they are first thermally excited to an energy between $E_F$ and the top of the barrier and then tunnel to the metal (thermally-assisted tunneling).  The forward-bias $I-V$ characteristics in the direct tunneling regime are expressed as
\begin{eqnarray}
J = J_S\exp(qV/E_{\rm 00})
%\hspace{0.5cm} 
\label{eq:field}
\end{eqnarray}
where $J_S$ depends weakly on temperature as $cT/\sin(cT)$, where $c$ is a constant. $E_{\rm 00}$ is a temperature-independent parameter and is given by
\begin{eqnarray}
E_{\rm 00} = \frac{qh}{4\pi}\left[\frac{N_d}{m^*\varepsilon_s}\right]^{1/2}
%\hspace{0.5cm} 
\label{eq:E00}
\end{eqnarray}
where $h$ is Planck's constant, $N_d$ is the donor concentration, and $\varepsilon_s$ is the permittivity of the semiconductor. As Eq.~\ref{eq:field} shows, $E_{\rm 00}/q$ corresponds to the inverse of slope in ln$\ J - V$ plots, and $E_{\rm 00}$ = 15.6 meV for the oxygen deficient junction and 9.9 meV for the stoichiometric junction according to the slope observed at 10 K for $H=0$.
 
\begin{figure}
\begin{center}
\includegraphics[width=7.5cm]{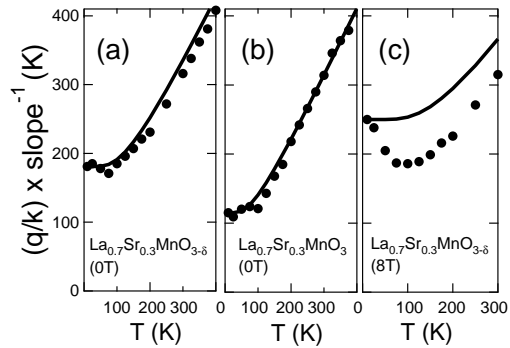}
     \caption{Inverse of slope in ln$J-V$ plots (dots) and $E_0/k$ following Eq.~\ref{eq:E0} (curves) for the oxygen deficient junction at 0 T (a), stoichiometric junction at 0 T (b), and oxygen deficient junction at 8 T (c).}
     \label{fig:slope}
\end{center}
\end{figure}

On the other hand, in the thermally-assisted tunneling regime the energy of tunneling electrons is distributed around $E_m$ and both $E_m$ and the distribution energy width increase as the temperature increases. There the $I-V$ characteristics are given by 
\begin{eqnarray}
J = J_S\exp(qV/E_{\bf 0}),
%\hspace{0.5cm} 
\label{eq:thermionic_field}
\end{eqnarray}
\begin{eqnarray}
E_{\rm 0} = E_{\rm 00}\coth(E_{\rm 00}/kT)
%\hspace{0.5cm} 
\label{eq:E0}
\end{eqnarray}
and
\begin{eqnarray}
J_S = \frac{A^*T^2\pi^{1/2}E_{\rm 00}^{1/2}\left[q(\phi_{bn} - V) + \xi\right]^{1/2}}{kT\cosh(E_{\rm 00}/kT)}\nonumber\\ \times \exp
\left[\frac{\xi}{kT} - \frac{q\phi_{bn} + \xi}{E_{\rm 0}}\right],
%\hspace{0.5cm} 
\label{eq:JS_cosh}
\end{eqnarray}
where  $\xi$ is the energy difference between $E_F$ and the conduction band bottom of the semiconductor. 
Again, $E_{\rm 0}/q$ corresponds to the inverse of  slope in  $\ln J - V$ plots. 
Equation~\ref{eq:E0} shows that $E_0$ approaches $E_{\rm 00}$ in the low temperature limit.  The thermionic emission regime is recovered at high temperature and specifically, slope$^{-1} \times (q/k)$ approaches the measurement temperature for $E_{\rm 00} \ll kT$. In Fig.~\ref{fig:slope} (a) and (b) we compare the inverse of  slope of ln $J-V$ plots at $H=0$ with $E_0/k$ calculated according to Eq.~\ref{eq:E0}, where $E_{\rm 00}$ is estimated from the observed slope at 10 K. The figure shows
 that for both junctions, the single-parameter ($E_{00}$) function of Eq.~\ref{eq:E0} accurately reproduces the observed slope up to 400 K, indicating that $[N_d /(m^* \varepsilon_s)]$ is  temperature independent between 10 and 400 K for $H=0$.
Equation~\ref{eq:E00} with $E_{\rm 00}$ = 15.6 meV and with 9.9 meV gives $m_r \varepsilon_r$ =$(m^*/m_0) \times (\varepsilon_s/\varepsilon_0)$ = 4.7 and 11.6, respectively, for $N_d = 3.32 \times 10^{18} {\rm cm}^{-3}$ which is the nominal carrier concentration of  Nb 0.01 weight \% doped SrTiO$_3$. 
These values, which are quite small for SrTiO$_3$, may reflect a significant decrease in
the permittivity of doped SrTiO$_3$  due to proximity to the interface \cite{Yoshida} or due to the large electric field in the junction structures.\cite{Christen}

\begin{figure}
\begin{center}
\includegraphics[width=7.5cm]{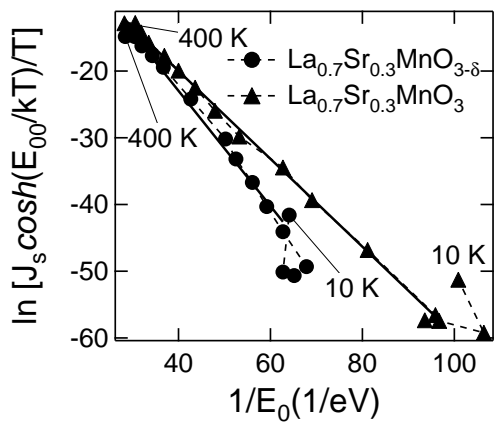}
     \caption{
     $\ln[J_S\cosh(E_{\rm 00}/kT)/T]$ plotted as a function of $1/E_{\rm 0}$  by dots for 
     the oxygen deficient junction and by triangles for the stoichiometric junction. The data points are connected by dashed lines. Bold lines are linear fits to the data between 400 K and 100 K.
          }
     \label{fig:cosh}
\end{center}
\end{figure}

 If $\xi$ is small compared with $q\phi_{bn}$, the slope of  $\ln[J_S\cosh(E_{\rm 00}/kT)/T]$ versus $1/E_{\rm 0}$ plot corresponds to $- q\phi_{bn}$ according to Eq.~\ref{eq:JS_cosh}. As shown in Fig.~\ref{fig:cosh}, such plots for the junctions studied here at $H=0$ are almost linear above $\sim$ 100 K and we find $\phi_{bn} \sim$ 0.86 V for the oxygen deficient junction and $\sim$ 0.66 V for the stoichiometric junction, meaning that the gradual decrease in $\phi_{bn}$ in going from 400 K to 100 K in Fig.~\ref{fig:JS} reflects a change in the energy of the thermal assistance rather than a change in the barrier height. The result that both the barrier height and $[N_d /(m^* \varepsilon_s)]$ are independent of the temperature is striking since the capacitance of both junctions shows a temperature dependence of a factor two.\cite{Nakagawa2} 

Contrasted to the junction characteristics at $H=0$, which are very similar to those of conventional metal-semiconductor Schottky junctions, the characteristics for the oxygen deficient junction under magnetic field are quite unusual, as shown in Fig.~\ref{fig:IV} (c): as the temperature decreases the slope first increases between 300 K and 150 K, shows almost no 
temperature dependence around 100 K, and finally decreases again below 50 K, as visible in panel (d) of the figure. In Fig.~\ref{fig:slope} (c) we compare the inverse slope and calculated $E_0/k$, where $E_{\rm 00}$  has again been determined from the observed slope at 10 K. Since direct tunneling without thermal excitation gives $d$(slope$^{-1}$)/$dT$ = 0, and both thermally-assisted tunneling and thermionic emission processes  give $d$(slope$^{-1}$)/$dT > 0$ for all temperature, the temperature dependence of $d$(slope$^{-1}$)/$dT < 0$ observed below 100 K is incompatible with the temperature-independent $E_{\rm 00}$. By contrast, the $I-V$ characteristics do not change by applying the magnetic field for the oxygen stoichiometric junction.

Given the strong quantitative agreement between the observed data and Schottky picture for $H=0$, as shown in Fig.~\ref{fig:slope} (a)(b) and Fig.~\ref{fig:cosh}, mechanisms within the framework of the Schottky picture, if any, would be relevant also for
the unusual temperature dependence of $d$(slope$^{-1}$)/$dT < 0$, although the present result does not exclude the possibility that localized charge in interface states play some role.\cite{Sze} 
In the context of this analysis, the magnetic field changes the effective Schottky barrier for junction transport, in a temperature dependent manner.  In order to understand the implications of this result, we emphasize that there is no spin selector at the interface, as there is in CMR (neighboring spin orientations), or TMR (relative electrode or grain magnetization orientations).  Furthermore, in the experiments presented here, all the potential drop is across the interface itself, with negligible contributions across the manganite film.  Therefore, the magnetic field response does not reflect the magnetoresistance of the film.

Based on the Schottky picture, the temperature dependence of $d$(slope$^{-1}$)/$dT < 0$ should have two origins. One is recurrent temperature dependence of $\varepsilon_s$ in titanate, which gives temperature dependence to $E_{00}$: the magnetic field shifts the chemical potential of manganite \cite{Wu} and hence modifies the barrier height, resulting in a change in the electric field in titanate. Such a mechanism would turn on the temperature dependence of $E_{00}$ since the presence or absence of the temperature dependence in $\varepsilon_s$ of SrTiO$_3$ is a function of the electric field. 
The second origin is the effect of the unoccupied density of states (DOS) of manganite. In the direct tunneling regime, the final-state unoccupied DOS of manganite is involved in the forward-bias $I-V$ characteristics in an equal manner as the initial-state DOS. A recent O K-edge x-ray absorption measurement of La$_{1-x}A_x$MnO$_3$ ($A$ =Ca, Sr) \cite{Mannella} shows that the double-peak structure just above $E_F$ is smeared as the temperature increases. Such a non-ideal DOS that depends on magnetic field and on temperature leads to  deviation from the simple Schottky picture. 
%although why such effect appears only for the oxygen deficient junction under the magnetic field is again unclear.
These two mechanisms are very different from either CMR or TMR, where the spin-dependent DOS in the vicinity of $E_F$ play a dominant role. 
 
 In conclusion,  the $I-V$ characteristics of La$_{0.7}$Sr$_{0.3}$MnO$_{3(-\delta)}$/Nb:SrTiO$_3$ junctions are found to be consistent with (thermally-assisted) tunneling across the Schottky barrier at $H=0$, while those of magnetoresistive junctions significantly deviate from Schottky behavior under magnetic field. Such $I-V$ characteristics reflect changes in the chemical potential and in the unoccupied DOS of manganite, both of which have not intensively been studied nor utilized for device applications thus far.

We thank A. F. Hebard, Y. Kozuka, and Y. Hikita for discussions. N.N. acknowledges partial support from QPEC, Graduate School of Engineering, University of Tokyo. 

%\renewcommand{\baselinestretch}{2.5}

%\begin{figure}
%     \caption{Temperature dependence of the forward-bias $I-V$ characteristics for La$_{0.7}$Sr$_{0.3}$MnO$_{3-\delta}$ junction under the magnetic field of 8 T (a). We enlarge the low-temperature characteristics in (b). Bold lines again correspond to linear fitting on semilog scale.
%          }
%     \label{fig:IV_mag}
%\end{figure}
%
%\begin{figure}
%     \caption{Dots: 
%     Inverse of slope in ln$J-V$ plots under no magnetic field for La$_{0.7}$Sr$_{0.3}$MnO$_{3-\delta}$ junction at 8 T (dots) compared with $E_0/k$ (curve).}
%          \label{fig:slope_mag}
%\end{figure}

%\end{multicols}

\end{document}